\documentclass{article} 
\usepackage{nips15submit_e,times}
\usepackage{hyperref}
\usepackage{url}
\usepackage{graphicx}
\usepackage{amsmath}
\usepackage{amssymb}
\usepackage{subfig}

\title{Network Formation Game with Different Cost Value for Players}
\author{
Dinesh Govindaraj \\
Department of Computer Science and Automation \\
Indian Institute of Science \\
Bangalore, India }

\nipsfinalcopy 
\begin{document}
\maketitle

\begin{abstract}
In Social Networks, it is often interesting to study type of networks formed, its efficiency with respect to social objective and which networks are stable. Many work have already been there in this area. Players in network formation game will incur some cost to form a link. In previous work, all player will have same cost for forming a link int he network. Here we will the study the optimal network and Nash equilibrium network when each player have different cost value for forming the link in the social network.
\end{abstract}

\section{Introduction}

A social network is a social structure made of nodes (which are generally individuals or organizations). Nodes are tied by one or more specific types of interdependency, such as values, visions, ideas, financial, exchange, friendship, kin- ship, dislike, conflict or trade. The resulting graph-based structures are often very complex. Problems in Social Net- works are 1) Type of networks formed and its efficiency with respect to social objective, 2) Diffusion of innovations theory explores social networks and their role in influencing the spread of new ideas and practices and 3) Searching in Social Networks. In network formation game, the Players are nodes, and their strategy choices create an undirected graph. Each node chooses a subset of the other nodes, and lays down edges to them. The edges are undirected, in that, once in- stalled, they can be used in both directions, independently of which node paid for the installation. The union of these sets of edges is the resulting graph.

\section{Previous Work}
Often, we need to find how inefficient is the network formed. Some of the measures of inefficiency \cite{game3,game4} of equilibrium are, Price of anarchy and Price of Stability. Price of anarchy is defined as the ratio between the worst objective function value of an equilibrium of the game and that of an optimal outcome. Price of Stability is defined as the ratio between the best objective function value of one of its equilibria and that of an optimal outcome. In \cite{game1,game2}, authors proposed the following model for the network formation game and derived bounds on Price of anarchy. 

The Model is described as follows. Player $[n] = \{0,1,2,...,n-1\}$, Strategy Space $S_i = 2^{[n]-\{i\}}$ and Strategy Profile $s = \{s_0,s_1,s_2,...,s_{n-1}\}$ where $s_i \in S_i$. Graph formed in this setting is $G[s] = \{[n], \cup_{i=0}^{n-1} ({i} \times s_i)\}$. Cost for each player in this game is $c_i(s) = \alpha.|s_i| + \sum_{j=0}^{n-1} d_{G(s)} (i,j)$, where $\alpha$ is cost for an edge and $d_{G(s)} (i,j)$ is distance between $i,j$ in G.

\section{The New Model}
In this new model, cost for forming the link by player $i$ is $\alpha_i$. That is, each user have different cost value to build the link. Cost for each player in this setting is $c_i(s) = \alpha_i |s_i| + \sum_{j=0}^{n-1} d_{G(s)} (i,j)$, where $d_{G(s)} (i,j)$ is distance between $i,j$ in G.

\section{Nash Equilibrium}
With out loss of generality consider $\alpha_i \le \alpha_2 \le ... \le \alpha_n$. The Nash equilibrium with different condition on values of alpha is as follows. \\ \\
\textbf{Case 1: $\alpha_n \le 1$}\\
If $\alpha_n \le 1$, then complete graph is a Nash equilibrium. Since $\alpha_n \le 1$, if any node tries to remove an edge, distance will increase by 1 but cost for forming link is less or equal to 1. \\ \\
\textbf{ Case 2: $\alpha_1 > 1$ } \\
If $\alpha_1 > 1$, then star graph is a Nash equilibrium. Since $\alpha_1 > 1$, if any node tries to add an edge, distance will decrease by 1 but cost for forming link is greater than 1. \\ \\
\textbf{Case 3:} General Case\\
Consider a node $j$ such that $\alpha_j \le 1$ and $\alpha_{j+1} > 1$. If there exist node $j$ such that $\alpha_j \le 1$ and $\alpha_{j+1} > 1$, then Nash equilibrium graph will contain clique of size $j$ and $n-j$ nodes connected to first $j$ nodes. \\ \\

\section{Social Cost}
Social Cost of the network in this will be $$C(G) = \sum_i c_i = \sum_{j=1}^n \alpha_j|s_j| + \sum_{u,v} d_G(u,v)$$. One lower bound on social cost is $$C(G) \le.|E| + 2|E| + 2(n(n-1)-2|E|)$$ $$C(G) \le 2n(n-1) + (\alpha_1 - 2) |E|$$
This is using the fact that there will $2|E|$ pairs of node will have dsitance one and all other pairs of node will have distance atleast two. If there is and edge $(i,j)$ in optimal network, then $i<j$, then $C_e(i,j) = \alpha_i$. \\ \\ 
\textbf{Case 1: $\alpha_1 > 2$}\\
$C(G) \geq 2n(n-1) + (\alpha_1 - 2) |E|$. $|E|$ should be minimum to minimize the social cost. So star graph is the social optimal solution because $|E|$ is $n-1$ which is minimum in any connected graph and distance function is minimum in the star graph. \\ \\
\textbf{Case 2: $\alpha_1 \le 2$}\\
$C(G) \geq 2n(n-1) + (\alpha_1 - 2) |E|$. $|E|$ should be maximum to minimize the social cost. So node $1$ should form maximum number of links. Lower bound on social cost in terms of node 2 given 
$\alpha_1 < 2$ is $$C(G) \geq \alpha_1.(n-1) + \alpha_2 (|E|-n+1) + 2|E| + 2(n(n-1)-2|E|)$$ $$C(G) \geq \alpha_1(n-1) + (\alpha_2 -2)|E| - (n-1)(\alpha_2-2n)$$ 
\textbf{Case 3: $\alpha_2 < 2$}\\
$C(G) \geq \alpha_1(n-1) + (\alpha_2 - 2) |E| - (n-1)(\alpha_2 -2n)$. $|E|$ should be maximum to minimize the social cost. So node 1 and node 2 should form maximum number of links.\\ \\
\textbf{Case 4: $\alpha_n < 2$}\\
Social optimal network is coplete network. Because $|E|$ should be maximum to minimize the social cost. So complete graph contains maximum number of edges and distance is also minimum in the complete graph. \\ \\ 
\textbf{Case 5: } General Case\\
If there exist node $j$ such that $\alpha_j \le 2$ and $\alpha_{j+1} > 2$, then social optimal graph will contain clique of size $j$ and $n-j$ nodes connected to first $j$ nodes. 

\section {Future Work}
We will come up with the lower bound on price of anarchy for general case. We will apply Myerson Value o this scenario in order to get cost for paying for the link. This will be similar to the Bargaining in networks and Myerson paper. 

{\small
\bibliographystyle{ieee}
\bibliography{egbib}
}

\end{document}